\documentclass[prb,preprint]{revtex4-1}
\usepackage{amsmath} 
\usepackage{graphicx}
\usepackage{float}

\begin{document}

\begin{center}
\Large\textbf{Leveraging AI for Rapid Generation of Physics Simulations in Education: Building Your Own Virtual Lab}

\vspace{1cm}
\normalsize
Yossi Ben-Zion$^1$\textsuperscript{*}, Roi Einhorn Zarzecki$^1$, Joshua Glazer$^1$, Noah D. Finkelstein$^2$

\vspace{0.5cm}
\small
$^1$Department of Physics, Bar-Ilan University, Ramat Gan IL52900, Israel\\
$^2$Department of Physics, University of Colorado Boulder, Boulder, Colorado 80309, USA\\
*benzioy@biu.ac.il
\end{center}

\vspace{1cm}

Seemingly we are not so far from Star Trek's food replicator. 

Generative artificial intelligence (AI) is rapidly becoming an integral part of both science and education, offering not only automation of processes but also the dynamic creation of complex, personalized content for educational purposes. With such advancement, educators are now crafting exams, building tutors, creating writing partners for students, and developing an array of other powerful tools for supporting our educational practices and student learning \cite{bahroun2023transforming,kortemeyer2024ethel,avila2024using,kuchemann2024large,steinert2024using}.  We share a new class of opportunities for supporting learners and educators through the development of AI-generated simulations of physical phenomena and models.  While we are not at the stage of "Computer: make me a mathematical simulation depicting the quantum wave functions of electrons in the hydrogen atom," we are not far off. 

Educators have long known and demonstrated the value of computer simulations for supporting student learning\cite{finkelstein2006high,redish1993student,christian2000physlets}. Simulations can complement text, formula, videos, physical demonstrations, and other educational media by focusing on key pedagogical features, such as: promoting engagement through interactivity, simplification of phenomena, emphasizing particular models, supporting play and "messing about", manipulating key physical parameters (even to the non-real regime, e.g., stopping time), and promoting student reflection and metacognition.  Their capacities for supporting student learning has been remarkably productive across a wide array of audiences,  from early school age through graduate students. 

At the same time, the development of simulations historically has come with a significant overhead and start-up cost, typically requiring coding expertise (or limiting to modifications of existing code) and software platforms to construct these simulations.  Furthermore, typically the content area, representations used in the simulation, and forms of interaction are decided by the developers, in advance of the educators and learners engaging with the software tools.  Such an approach has great strengths and weaknesses. Simulations can represent physically accurate models, embed pedagogically effective approaches, and deploy proven user-interface models.  On the other hand, simulations can be expensive, time-consuming and do not involve the end-user in the design.  The use of AI-generate simulations presented here can complement (not replace) these more robust and thoroughly developed simulations and broaden the space of tools used by learners and educators alike.

Using AI models such as ChatGPT (OpenAI)\cite{ChatGPT} and Claude (Anthropic) \cite{Claude} through predefined prompts, this approach democratizes simulation creation, enabling educators and students to design custom simulations without programming expertise. We illustrate this methodology through the development of simulations for foundational systems such as the simple pendulum, the Ising model, and the random walker, detailing implementation methods. This approach enhances experiential learning by fostering critical thinking and problem-solving skills while expanding access to tailored educational tools. Moreover, the process of validation and iterative refinement inherent in this method promotes valuable reflective and meta-cognitive activities, making it a powerful complement to traditional methods in physics education while supporting broader integration of interactive learning tools.

\section{Prompt-Based Simulations}

Using large language models (LLMs) we designed a prompt structure that is flexible enough to be customized for different physical systems, such as basic Mechanics, Electricity and Magnetism, or other more advanced dynamics.
Below is the template we developed for creating physics simulations.

\textit{You are a web developer tasked with creating an interactive physics simulation for students. Write an HTML file that simulates a generic physical system with the following features:
A real-time animation of the system.
Sliders to adjust system parameters such as mass, length, and initial conditions.
Buttons to start, pause, and reset the simulation.
Graphs that display relevant quantities (e.g., position, velocity, energy) as a function of time.
Ensure the simulation is responsive and works across different devices.}

After receiving the initial output from the AI, we engaged in a dialogue with the model to refine and tailor the simulations to our specific needs, as will be clarified in the following sections. Through this iterative process, we achieved high-quality simulations that can be customized for different educational contexts.

The implementation process is straightforward and requires no advanced programming knowledge: copy the AI-generated HTML and JavaScript code into a text editor, save it with an .html extension, and open it in any modern web browser to interact with the simulation through adjustable sliders and buttons. This approach enables educators to focus on teaching physics while students gain hands-on experience with interactive simulations.

\subsection{Validation of Results}

The accuracy and reliability of simulations created with Large Language Models (LLMs) are critical, yet challenging to ensure due to the inherent variability of these models. Even with identical prompts, LLMs can produce different outputs, stemming from their probabilistic nature. This variability necessitates comprehensive validation strategies.
 
We identify two main types of validation: \textbf{technical validation} and \textbf{physical validation}. Technical validation tests the simulation's behavior under various conditions and assess its response to parameter changes. Physical validation compares simulation results to known analytical solutions and verify consistency with established physical laws.

The validation process, particularly when performed by students, offers significant educational benefits. Students strengthen their physics understanding by examining simulation results against established laws and observing how changes in parameters affect the system's behavior. The iterative nature of working with LLMs—refining prompts, correcting errors, and re-running simulations—mirrors scientific inquiry and develops systematic problem-solving skills. Moreover, this exposure to cutting-edge AI technology provides valuable experience for their future learning and careers.

Importantly, these validations can be conducted through natural language dialogue with the model, requiring no additional programming expertise. Users can question, seek clarification, and explore various aspects of the simulation directly with the model, making the validation process accessible to all users.

\subsection{Language Models in Simulation}

When generating educational physics simulations, the choice of AI language model can significantly affect the validation process. Current leading models excel in different areas: some are particularly effective at breaking down complex tasks into manageable steps and handling detailed prompts with precision, while others offer advantages in speed and real-time adaptability, including immediate HTML compiling and display. These capabilities enable both educators and students to create effective interactive simulations. While these models continue to evolve rapidly, they consistently demonstrate the ability to generate valid physics simulations, though some corrections through human supervision typically remain necessary\footnote{As of 2024, the leading models are OpenAI's O1 and Claude 3.5 Sonnet. O1 exemplifies the detailed step-by-step approach, while Claude 3.5 Sonnet demonstrates the rapid adaptability features. GPT-4o can also be effective for simpler simulations with shorter prompts. The choice between them depends on specific project requirements.}.

In the following sections, we present the process of building simulations for three basic physical systems commonly used in physics education: the simple pendulum, the Ising model, and the random walker. These systems were selected as they represent distinct cases: a system governed by differential equations (the simple pendulum), a many-body statistical system (the Ising model), and a simple stochastic process (the random walker). The method we demonstrate can be extended to a wide variety of physical systems, but these examples were chosen to showcase the breadth of applications.

In all three cases, the simulations can provide substantial added value in understanding the systems, particularly in the cases of the Ising model and random walker, where simulations are the most common method of demonstration due to the significant challenges in constructing simple experiments or demonstrations. For each of these models, we will present the prompt used to generate the simulation and provide details on possible methods for validating the results.

\section{Simple Pendulum}

The simple pendulum is a fundamental system in classical mechanics, often used to demonstrate harmonic motion and to explore the limits of the small-angle approximation. While the analytical solution for small angles is straightforward, capturing the full non-linear dynamics for larger displacements requires numerical methods.

\subsection{Objective of the Simulation}
This simulation provides an interactive tool for exploring the dynamics of a simple pendulum. By adjusting parameters such as mass, pendulum length, initial angle, and angular velocity, students can observe their impact on the pendulum’s motion. This allows users to explore relationships such as how the pendulum length affects the period of the pendulum and how larger initial angles deviate from the small-angle approximation.

\begin{figure}[h!]
    \centering
    \includegraphics[width=\textwidth]{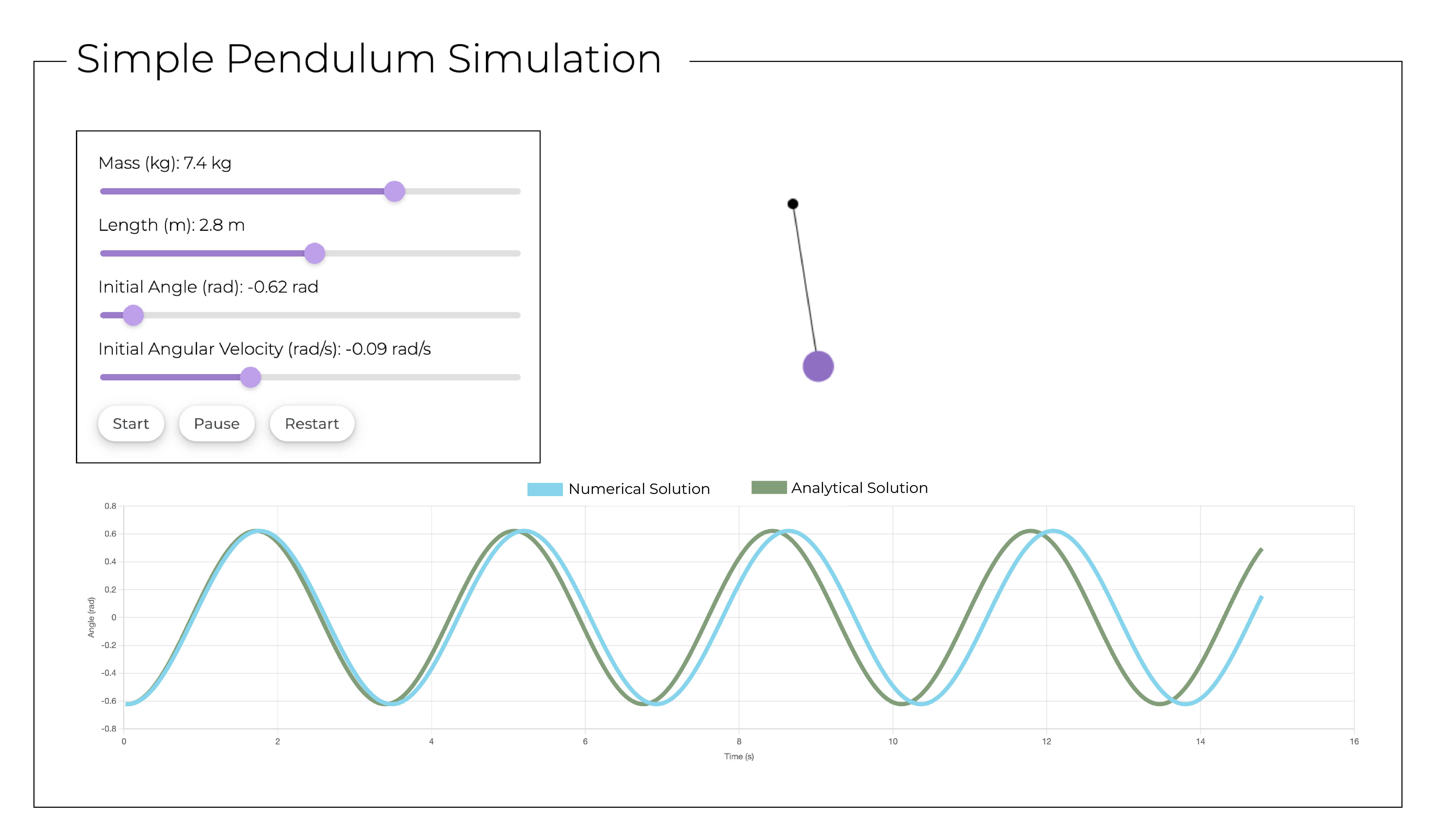}
    \caption{Basic pendulum simulation interface showing the pendulum motion, parameter controls, and time-dependent graphs of angle and angular velocity.}
    \label{fig:simple_pendulum_simulation}
\end{figure}

As shown in Figure~\ref{fig:simple_pendulum_simulation}, the simulation includes a visual representation of the pendulum's motion with adjustable parameter sliders. The accompanying graphs display the pendulum's angle and angular velocity over time, enabling dynamic exploration of the system's behavior.

\begin{figure}[H]
    \centering
    \includegraphics[width=\textwidth]{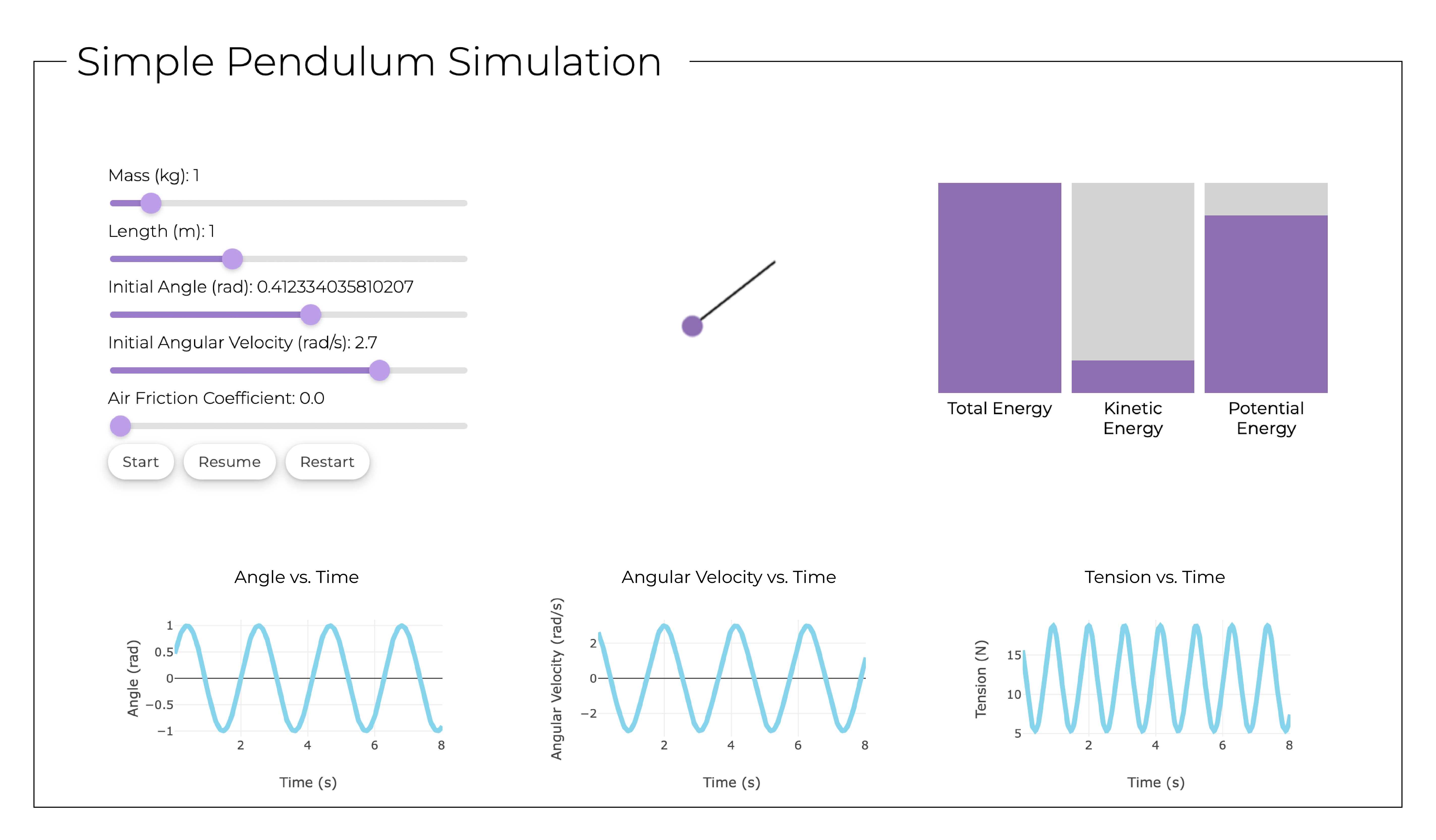}
   \caption{Enhanced pendulum simulation featuring large-angle dynamics, friction effects, pendulum tension graph, and energy distribution visualization.}
    \label{fig:advanced_pendulum_simulation}
\end{figure}

An enhanced version of the pendulum simulation (see Figure~\ref{fig:advanced_pendulum_simulation}) explores behavior beyond simple harmonic motion, including full rotations and non-periodic trajectories. A friction slider demonstrates damping effects on the motion, while additional visualizations display pendulum tension over time and energy distribution through bars representing total, kinetic, and potential energy. These features provide students with deeper insights into the complex dynamics of pendulum motion.

\subsection{Original Prompt and Model Implementation}
The following prompt was used to generate the pendulum simulation with either OpenAI's O1 or Claude 3.5 Sonnet language models:

\textit{You are a web developer tasked with creating an interactive physics simulation for students. Write an HTML file that simulates a simple pendulum without friction. \\
Animation:\\
Display a real-time animation of the pendulum swinging on the page. \\
Controls: \\
Sliders to adjust the following parameters: \\
Mass (kg): Range from 0.1 kg to 10 kg, default 1 kg. \\
Length (m): Range from 0.5 m to 5 m, default 2 m. \\ Initial Angle (rad): Range from -0.7 to 0.7 radians, default 0.05 radians. \\
Initial Angular Velocity (rad/s): Range from -0.3 to 0.3 rad/s, default 0 rad/s.\\ Ensure each slider has a label and displays the current value. \\
Graph: \\
Angle vs. Time Graph: - Plot the pendulum’s angle over time. - Overlay the analytical solution for small angles ($theta(t) = theta0\cdot cos(t\cdot sqrt(g/L)) + (omega0/sqrt(g/L)) sin(t\cdot sqrt(g/L)))$ for comparison. \\
Display numerical values (ticks) on the graph axes.\\  
Include a legend for the numerical and analytical lines on the graph. \\
Simulation Controls: \\
Start Button: Begins the simulation. \\
Pause Button: Pauses/resumes the simulation without resetting. \\
Restart Button: Stops the simulation and resets to the initial conditions. \\
Layout: \\
Organize all elements neatly on the page. 
Ensure the simulation is usable on various screen sizes (responsive design). \\
Technical Requirements: \\
Use appropriate Runge-Kutta methods to solve the nonlinear equations of motion of the pendulum. Comment your code to explain key sections for educational purposes. Ensure all elements fit neatly on the page . Immediately update the drawing whenever any slider is adjusted. Reset the analytical and numerical graphs based on the new initial conditions whenever any slider is adjusted.}

This prompt generates a functional simulation with adjustable parameters and allows real-time visualization and comparison between the numerical solution and the analytical small-angle approximation.

\subsection{Testing Procedure}
To ensure the simulation is technically accurate and pedagogically effective, we conducted a series of tests. These tests are divided into two categories: technical tests and physical tests, each verifying different aspects of the simulation.

\subsubsection{Technical Tests}
 \begin{enumerate}
    \item Slider Interactivity: Ensure that adjusting the mass, length, initial angle, and angular velocity sliders updates the pendulum’s animation and graph in real-time.
    \item Axis Reset on Parameter Change: Verify that the graph resets and recalibrates when any of the sliders are adjusted, ensuring that the simulation reflects the new initial conditions accurately.
    \item Graphical Representation: Ensure the graph displays numerical values on both axes and includes a clear legend differentiating between the numerical and analytical solutions.
    \item Pendulum Length at Maximum Extension: Ensure the entire pendulum, including the pendulum, remains visible within the canvas at its maximum length.
\end{enumerate}

\subsubsection{Physical Tests}
\begin{enumerate} 
\item Period Accuracy for Small Angles: Confirm that for small initial angles, the period of the pendulum matches the analytical solution  $T = 2\pi \sqrt{\frac{L}{g}}$.
    \item Numerical vs. Analytical Solution for Larger Angles: Ensure that for small initial angles, the numerical and analytical solutions match, but as the initial angle increases, the numerical solution diverges from the analytical one, illustrating the limits of the small-angle approximation.
\end{enumerate}
Notably, this physical test becomes a great opporutnity for students to review and refine their own understanding, as they validate the output of the AI generated simulator.

If any issues arise during testing, we can fix them through re-prompting the AI. For example, if the sliders do not update the animation in real-time, we would send the prompt: "The pendulum animation currently doesn't respond immediately when I move the sliders. Please modify the code to make it update in real-time." The AI will then provide the corrected code that addresses this specific issue.

To illustrate our methodology in practice and highlight the iterative nature of working with LLMs, we present a debugging session using Claude 3.5 Sonnet. The following prompts were used to refine the simulation: 
\textit{"Reset the analytical and numerical graphs based on the new initial conditions whenever any slider is adjusted.", 
"Update the graphs throughout the entire simulation.", 
"Change the pendulum length - rescale to $50\%$ of its current length.", 
"Fix the analytical graph. There is a problem with the analytical solution shown - it displays a period that is exactly half of the correct time. Check if you used the function I gave you in the prompt. Check if the time in the analytical solution is correct."}
We found that effective prompts combine clear problem identification with specific correction instructions. This approach proved most successful in obtaining accurate responses from the LLM, demonstrating the importance of precise communication in the refinement process.

\section{Ising Model}

The Ising model is a crucial framework in statistical physics, primarily used to study ferromagnetism and phase transitions in materials. In its simplest form, it consists of a lattice of spins that interact with their nearest neighbors, where each spin can be in one of two states (up or down). The energy of the system depends on the alignment of these spins, with neighboring spins that are aligned reducing the system's energy, while misaligned spins increase it. The Ising model is particularly valuable for studying how microscopic interactions at the local level give rise to macroscopic phenomena, such as the transition between magnetized and non-magnetized states as temperature varies. Beyond physics, the model has been applied in areas such as biology and social sciences, making it a versatile tool for exploring collective behavior in complex systems.

\subsection{Objective of the Simulation}
\begin{figure}[h!]
    \centering
    \includegraphics[width=\textwidth]{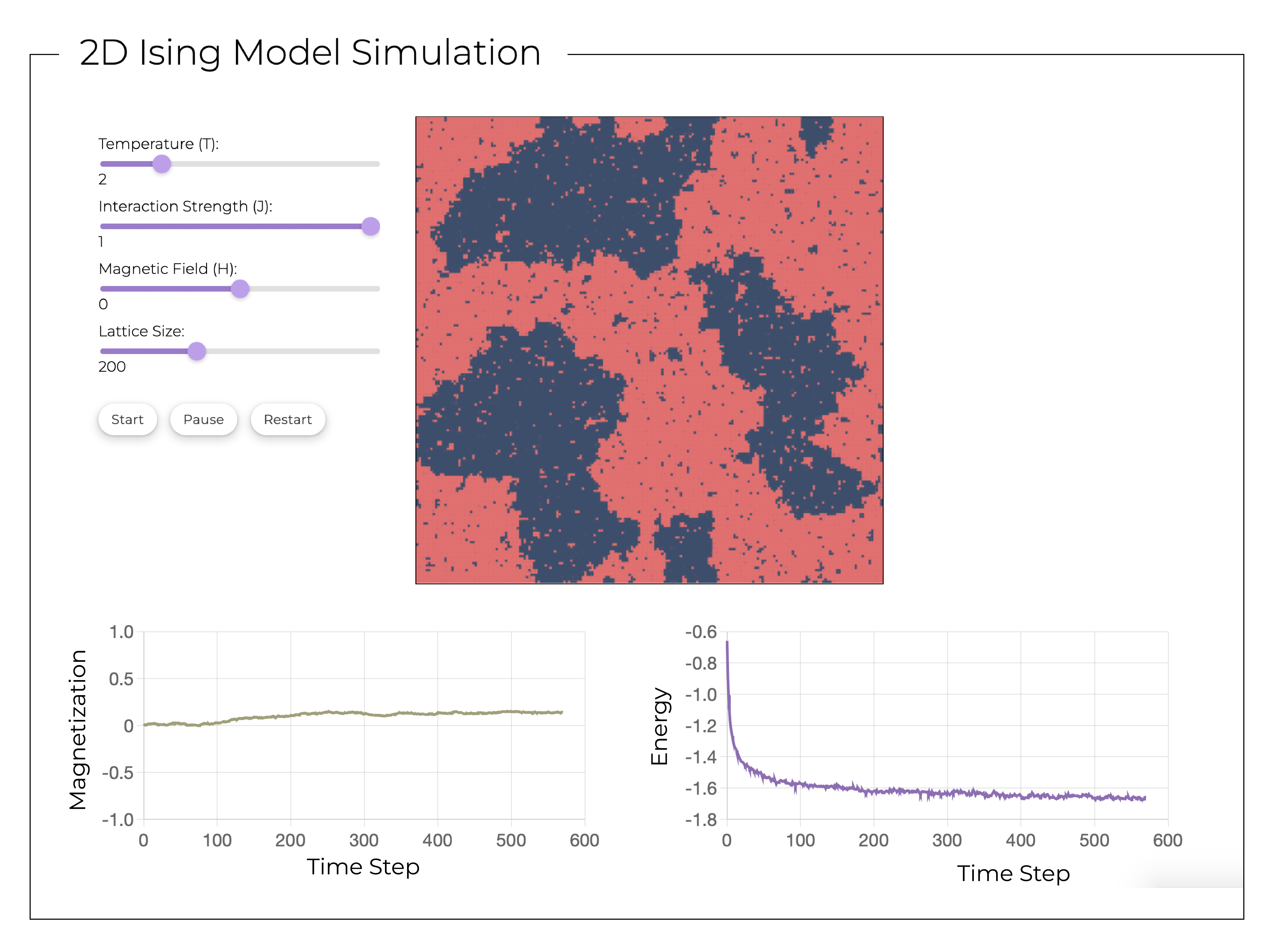}
    \caption{2D Ising model simulation with heat map visualization, parameter controls, and real-time graphs of magnetization and energy.}
    \label{fig:ising_model_simulation}
\end{figure}

The goal of this simulation is to provide an interactive tool for exploring the behavior of the 2D Ising model, particularly its phase transitions and magnetization properties. By allowing users to adjust parameters such as temperature, interaction strength, magnetic field, and lattice size, the simulation offers students a hands-on opportunity to observe how local spin interactions lead to global changes in system order. It also allows users to explore the critical temperature, where the system undergoes a phase transition from a magnetized (ordered) state to a disordered state. To visualize the dynamic behavior of the Ising model, we included a simulation (see Figure~\ref{fig:ising_model_simulation}) that represents the spin orientations on a 2D lattice through a heat map, while real-time graphs display the system's magnetization and total energy over time, providing insight into phase transitions and response to external magnetic fields.

\subsection{Original Prompt and Model Implementation}
The following prompt was used to generate the Ising model simulation using OpenAI’s O1 and Claude 3.5 Sonnet models:

\textit{"You are a web developer tasked with creating an interactive physics simulation for students. Write an HTML file that simulates the 2D Ising model. The system should initialize with a random spin configuration and display a real-time heat map representing the spin orientations with periodic boundary conditions. Provide user controls through sliders for Temperature (T: 0 to 10, default 2.5), Interaction Strength (J: -1 to 1, default 1), Magnetic Field (H: -5 to 5, default 0), and Lattice Size (50x50 to 500x500, default 100x100). Include real-time graphs for Magnetization vs. Time and Total Energy vs. Time. The simulation should have start, pause, and reset controls. Use Monte Carlo methods (Metropolis-Hastings) for the simulation and ensure responsive updates."}

\subsection{Testing Procedure}
\subsubsection{Technical Tests}
As with the simple pendulum simulation, the Ising model simulation underwent technical checks to ensure correct functionality. This included verifying that the display of the spin orientations, the interaction sliders, and the real-time graphs were properly synchronized and responsive to changes in user inputs.

\subsubsection{Physical Tests}
The physical tests focused on ensuring that the simulation reflects the expected physical behaviors of the 2D Ising model under various conditions. Again, the particular areas of attention to physical validation could be the focus of an instructor prompting students, as they seek to focus on specific aspects of the Ising model and its applications  The main behaviors tested we included:
\begin{enumerate}
    \item Phase Transition and Critical Temperature: Near the critical temperature $T_c = 2.269$ for $J=1, H=0$, the system should undergo a transition from a magnetized (ordered) state to a disordered state. The magnetization should sharply decrease as the temperature approaches and exceeds this critical point. This behavior serves as a primary validation test for the simulation.
    \item Magnetization at Low Temperatures: At temperatures significantly lower than Tc, the system is expected to exhibit spontaneous magnetization, where most of the spins align in the same direction. This phenomenon should occur even without an external magnetic field. The simulation was tested to confirm this behavior by observing how magnetization increases as temperature decreases.
    \item Response to External Magnetic Field: When a magnetic field is applied $(H > 0)$, the spins should align with the direction of the field, causing an increase in magnetization proportional to the strength of the field. This test was performed at both high and low temperatures to ensure that the simulation reflects the expected physical response to external fields.
\end{enumerate}

\section{Random Walker} 
The random walker model is a key tool in understanding stochastic processes and statistical mechanics. In two dimensions, the walker takes random steps in any direction, making the model an ideal way to study diffusion, Brownian motion, and similar random phenomena. The model’s simplicity and relevance to real-world systems make it particularly useful for educational purposes, allowing students to grasp how random processes evolve over time. 

\subsection{Objective of the Simulation}
\begin{figure}[h!]
    \centering
    \includegraphics[width=\textwidth]{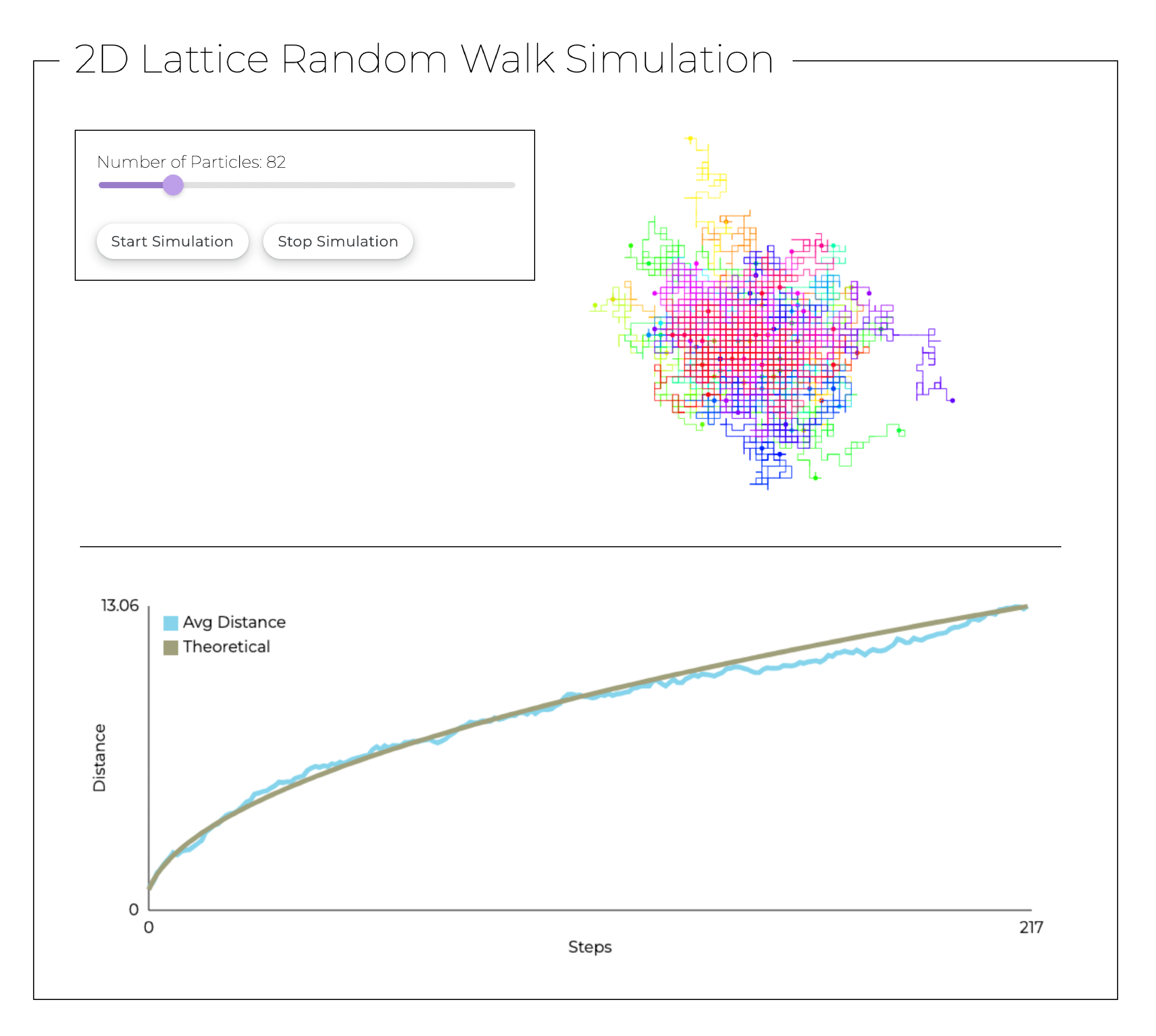}
        \caption{Multiple random walkers on a 2D lattice with adjustable walker count. The graph compares the average distance from origin with the theoretical prediction \(\langle r \rangle = \sqrt{\pi n / 4}\).}
    \label{fig:random_walker_simulation}
\end{figure}

The simulation provides a visual and interactive platform for exploring random walks. Students can adjust the number of walkers to observe how their average distance from the origin evolves compared to theoretical predictions. This hands-on approach helps demonstrate how individual random steps relate to broader phenomena like diffusion and molecular motion. Figure~\ref{fig:random_walker_simulation} shows the paths of multiple walkers and graphs their average distance from the origin over time, compared with theoretical expectations.

\subsection{Original Prompt and Model Implementation}
The following prompt was used to generate the simulation with OpenAI’s O1 and Claude 3.5 Sonnet models:\\
\textit{"You are a web developer, tasked with creating an interactive physics simulation for students. Write an HTML code to simulate a 2D lattice random walk of N points, with a step length of 1. Trace the path each point takes as they move around with a thin line. Add a graph that shows the average distance of all the points from the center as a function of time. Overlay the expected value, sqrt( pi * n / 4 ), on the same graph, where n is the number of steps. Add start, pause and restart buttons to the simulation. Add a slider to select how many points to simulate (min 1, max 500, default 5). Add numbers to the axes, titles for the axes, and a legend for each line on the graph."}

\subsection{Testing Procedure}
The technical validation verified the functionality of all interactive features, including simulation controls and the walker count slider, along with proper graph display elements.

The physical validation focused on comparing the walkers' average distance from origin with the theoretical prediction $\langle r\rangle =\sqrt{\frac{\pi n}{4}}$, confirming the model's consistency with 2D random walk analytics as the number of walkers increases.

\subsection{Random Walker - Expansion to 3D} 

\begin{figure}[h!]
    \centering
    \includegraphics[width=\textwidth]{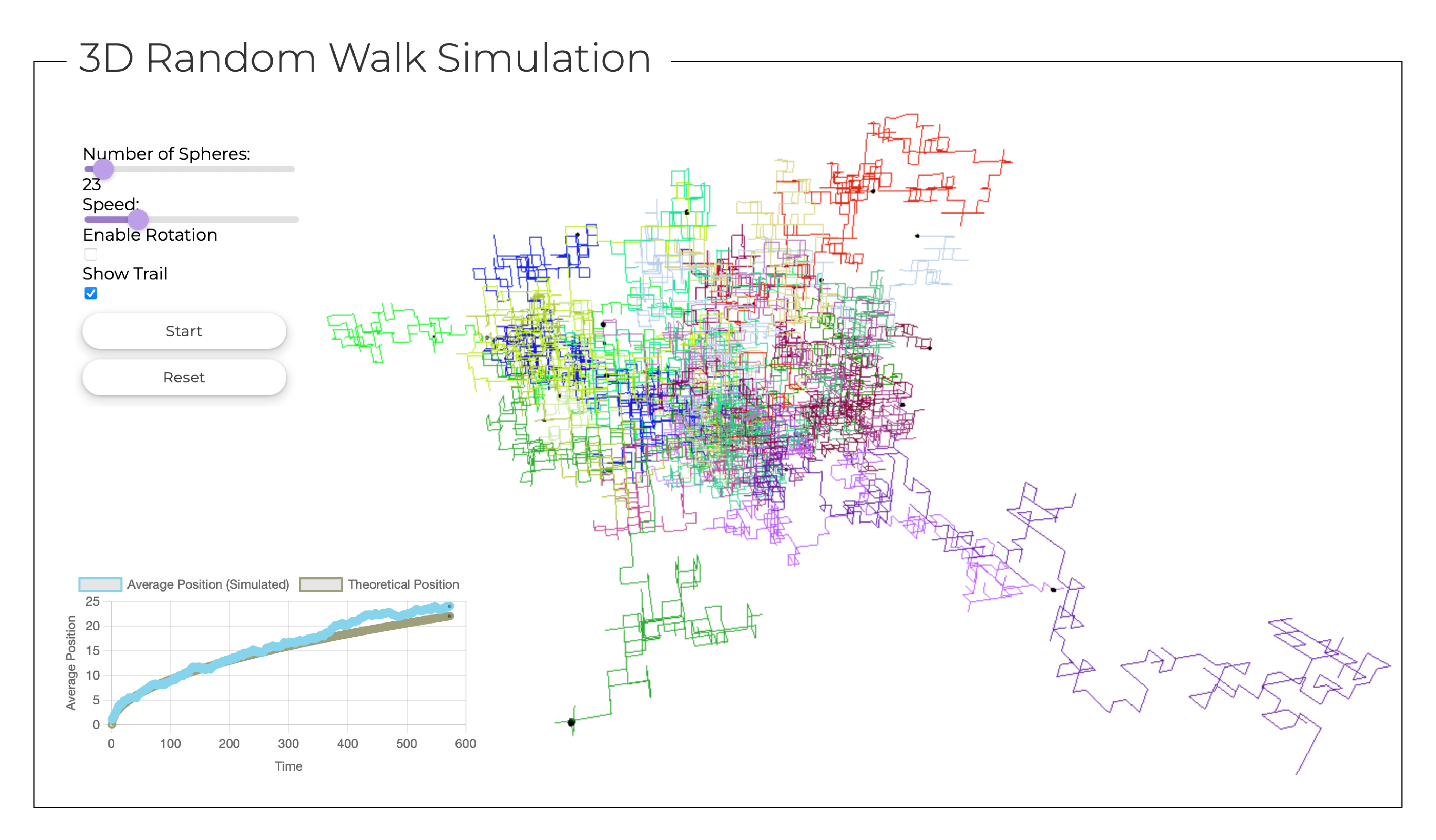}
    \caption{3D random walk simulation with adjustable particle count and viewing options. The graph compares average particle distance from the origin with the theoretical prediction \(\langle r \rangle = \sqrt{\frac{8}{3\pi} \cdot n}\).}
    \label{fig:3d_random_walker_simulation}
\end{figure}

The transition to a 3D simulation enables exploration of random processes in a fully spatial context, better representing particles moving in three dimensions. Figure~\ref{fig:3d_random_walker_simulation} shows particles moving in 3D space, with controls for particle count, rotation, trail visibility, and a graph comparing average distance from origin to theoretical predictions.

The initial prompt used with OpenAI's O1 was: \textit{"You are a web developer, tasked with creating an interactive physics simulation for students. Write an HTML code to simulate a 3D lattice random walk of N points, with a step length of 1. Trace the path each point takes as they move around with a thin line. Add a graph that shows the average distance of all the points from the center as a function of time. Overlay the expected value, sqrt(( 8 / 3pi) * n ), on the same graph, where n is the number of steps. Add start, pause and restart buttons to the simulation. Add a slider to select how many points to simulate (min 1, max 500, default 5). Add numbers to the axes, titles for the axes, and a legend for each line on the graph. Add controls for zooming and rotating the camera.}"

This prompt may lead to inconsistent results because of the task's complexity. Instead, we found it more effective to develop the simulation incrementally, adding features in stages ensuring that each stage worked correctly before introducing the next element. We started with a basic setup of a single red sphere performing a random walk in 3D space, then added functionality to control multiple particles via a slider. Path tracing was implemented next, allowing visualization of each particle's trajectory. Control buttons (Start, Pause, Reset) and display options (rotation and trail visibility) were then added to enhance user interaction. Finally, we incorporated statistical analysis with a graph showing the average distance from origin over time, compared with the theoretical curve ⟨r⟩ = sqrt((8/3pi)*n). Each stage was thoroughly tested before proceeding to the next, ensuring robust functionality throughout the development process.

This incremental approach ensured robust functionality and serves as a template for developing other 3D physics simulations.

\section{Conclusion}
This paper has presented a practical approach to generating educational simulations using large language models (LLMs), providing a user-friendly and highly accessible tool for lecturers and students alike. These tools enable the rapid creation of simulations tailored to specific educational needs, without requiring advanced programming skills, thus potentially significantly enhancing the teaching of complex physical science concepts by making them more accessible.

For educators, the ability to create custom simulations tailored to course material not only fosters an interactive learning environment, but also aligns with active learning principles, fostering student engagement. By integrating these tools into teacher training programs, educators are better equipped to design dynamic, responsive classrooms that encourage deeper conceptual understanding.

From the students' perspective, engaging in the creation of simulations promotes both deeper learning of the subject matter and hands-on experience with large language models. This process allows students to explore the strengths of LLMs in generating accurate simulations while recognizing their limitations, such as the need for careful prompt engineering and refinement. By experimenting with various scenarios, students develop critical thinking skills, learning to assess the reliability of these tools in scientific exploration. This approach fosters an active learning environment, increasing both motivation and engagement with the material.

It is important to emphasize that this approach is not intended to replace well-established, robust simulations like PhET, but rather to complement existing educational tools. This approach may be beneficial in cases where ready-made simulations are unavailable, when exploring edge cases not supported by existing tools, or as a comparative learning tool. The ability to both use professional simulations and create custom ones provides educators and students with a broader range of educational possibilities, enhancing the learning experience.

As LLMs continue to evolve, their potential for creating increasingly complex simulations will only grow. Advancements in natural language processing (NLP) and AI capabilities will make the process more intuitive, reducing the need for detailed prompt engineering. This evolution will allow educators and students alike to design simulations that more accurately mirror real-world systems, facilitating deeper exploration and analysis of complex phenomena.

While this paper has focused on applications in physics, the methods described are adaptable across disciplines, including chemistry, biology, and interdisciplinary studies. By applying domain-specific language models and refining prompts to suit the needs of various fields, this approach has the potential to significantly enrich learning experiences across a wide array of scientific disciplines, transforming the way students and educators engage with complex material.

\bibliographystyle{unsrt}
\bibliography{references}

\end{document}